\documentclass[useAMS,usenatbib]{mn2e}
\usepackage{graphicx,amssymb, epstopdf}
\newcommand{\vc}[1]{\bmath{#1}}

\newcommand{\beq}{\begin{equation}}
\newcommand{\eeq}{\end{equation}}
\newcommand{\bea}{\begin{eqnarray}}
\newcommand{\eea}{\end{eqnarray}}

\newcommand{\Secref}[1]{Section~\ref{#1}}

\newcommand{\eqref}[1]{equation~(\ref{#1})}

\newcommand{\lambdaeff}{\lambda_{\rmn{eff}}}
\newcommand{\vpar}{v_{\parallel}}

\title[Conduction in a mirror-unstable plasma]{Suppression of thermal conduction in a mirror-unstable plasma}
\author[S. V. Komarov, E. M. Churazov, M. W. Kunz and A. A. Schekochihin]{
S. V. Komarov$^{1,2}$\thanks{E-mail: komarov@mpa-garching.mpg.de}, E. M. Churazov$^{1,2}$, M. W. Kunz$^{3}$ and A. A. Schekochihin$^{4,5}$\\
$^{1}$Max Planck Institute for Astrophysics, Karl-Schwarzschild-Strasse 1, 
D-85741 Garching, Germany\\
$^{2}$Space Research Institute (IKI), Profsouznaya 84/32, Moscow 117997, Russia\\
$^{3}$Department of Astrophysical Sciences, Princeton University, Princeton, NJ 08544, USA\\
$^{4}$The Rudolf Peierls Centre for Theoretical Physics, University of Oxford, 
1 Keble Road, Oxford OX1 3NP, UK\\
$^{5}$Merton College, Oxford OX1 4JD, UK
} 
\begin{document}
\maketitle
\label{firstpage}

\begin{abstract}
The plasma in galaxy clusters is subject to firehose and mirror instabilities at scales of 
order the ion Larmor radius. The mirror instability generates 
fluctuations of magnetic-field strength $\delta B / B \sim 1$. These fluctuations act as magnetic 
traps for the heat-conducting electrons, suppressing their transport. We calculate 
the effective parallel thermal conductivity in the ICM in the presence of the mirror 
fluctuations for different stages of the evolution of the instability. The 
mirror fluctuations are limited in amplitude by the maximum and minimum values 
of the field strength, with no large deviations from the mean value. This key 
property leads to a finite suppression of thermal conduction at large 
scales. We find suppression down to $\approx 0.2$ of the Spitzer value for the 
secular phase of the perturbations' growth, and $\approx 0.3$ for their saturated phase. 
The effect operates in addition to other suppression mechanisms and independently of 
them. Globally, fluctuations $\delta B / B \sim 1$ can be present 
on much larger scales, of the order of the scale of turbulent motions. However, we 
do not expect large suppression of thermal conduction by these, because their scale 
is considerably larger than the 
collisional mean free path of the ICM electrons. The obtained suppression 
of thermal conduction
by a factor of $\sim 5$ appears to be characteristic and potentially universal 
for a weakly collisional mirror-unstable plasma.   
\end{abstract}

\begin{keywords}
conduction -- instabilities -- magnetic fields -- plasmas -- galaxies: clusters: intracluster medium 
\end{keywords}

\section{Introduction}
\label{sec:intro}

Thermal conduction in a magnetized plasma is a long-standing problem in 
astrophysics, dating back to the realization that virtually all astrophysical 
plasmas possess magnetic fields (based on both theoretical considerations 
and observations of synchrotron emission and the Faraday rotation). Although 
these fields are relatively weak ($\sim \ 1-10~ \rmn{\mu G}$  in the bulk 
of the intracluster medium (ICM), see, e.g., \citealt{Carilli2002} or \citealt{Feretti2012} for 
reviews), they constrain the motion of charged particles to spiraling 
along the field lines with Larmor radii typically very small compared 
to other physically relevant scales, namely, to the collisional mean free 
path and the correlation length of the plasma flows. In such a plasma, 
the electrons predominantly transfer heat along the field lines.

In the ICM, the quest for a theory of effective heat conductivity is 
strongly motivated by the observations of apparently long-lived temperature 
substructures \citep[e.g.,][]{Markevitch2003} and sharp gradients (cold 
fronts; e.g., \citealt{Markevitch2000, Ettori2000, Vikhlinin2001, 
MarkevitchVikhl2007}) that would not have survived had the electron 
conductivity been determined by the classic Spitzer expression for an 
unmagnetized plasma \citep{Spitzer1962}. Another puzzling topic is the 
stability of cluster cool cores, in which the role of thermal conduction 
is still unclear \citep[e.g.,][]{Rusz2002, Zakamska2003,VoigtFabian2004,Dennis2005}. 
  
The general problem of thermal conduction in an astrophysical plasma is 
greatly complicated by the fact that the medium is likely turbulent (for 
the ICM, see, e.g., \citealt{Inogamov2003, Schuecker2004, Schek2006turb, 
Subramanian2006, Zhur2014}), and so the magnetic-field lines are randomly 
tangled. It is practical to subdivide the problem into more narrowly 
formulated questions and study them separately. First, parallel conduction 
in a static magnetic field of a given structure can be investigated 
\citep[e.g.,][]{CC1998}. The static approximation is reasonable because 
electrons stream along magnetic fields faster than these fields are 
evolved by turbulence. Next, one can study the effective boost of the 
transverse conduction across the field lines due to their exponential 
divergence \citep{Skilling1974, RR1978, CC1998, Narayan2001, Malyshkin2001,
ChandranMaron2004}. Finally, local heat fluxes at the scale of turbulent 
eddies are affected by the correlation between temperature gradients and 
the magnetic field as they evolve in the same turbulent velocity field 
(\citealt{Komarov2014}; this process occurs on longer time scales than 
the other two). In this work, we only address the first part of the problem, 
parallel thermal conduction, as applied to the ICM. 

\begin{figure}
\includegraphics[width=84mm]{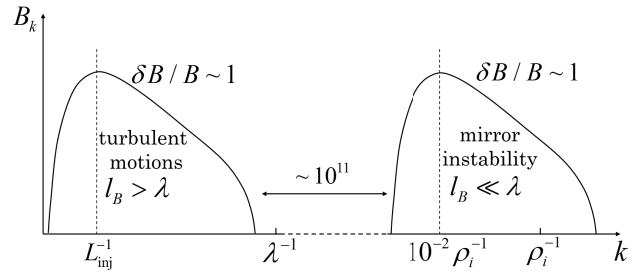}
\caption{A sketch of the spectrum of the fluctuations of magnetic-field strength  
in the ICM. The perturbations $\delta B/B \sim 1$ (relevant 
for magnetic trapping) generated by turbulence occupy the region 
$\lambda \lesssim l_B$, where magnetic trapping is ineffective. The 
mirror fluctuations, in contrast, are at the scales comparable to 
the ion Larmor radius $\rho_i,\ \lambda \sim 10^{13} \rho_i$, where 
magnetic mirrors can suppress electron transport considerably.} 
\label{fig:Bpar_sp}
\end{figure}

Parallel conduction can be affected by magnetic trapping of electrons by 
fluctuations of the field strength along a field line \citep{Ptuskin1995, 
CC1998, Chandran1999, MalyshkinKulsrud2001,Albright2001}. These fluctuations 
might be produced by various mechanisms. At the scale of turbulent motions, 
they can be generated by the small-scale turbulent MHD dynamo as a result 
of a series of random stretchings and compressions by the velocity field 
\citep[e.g.,][and references therein]{Schek2002,SchekCowleyTaylor2004,Schek2006turb}. At microscales 
of the order of the ion Larmor radius, the ICM plasma is subject to kinetic 
instabilities \citep{Schek2005, Schek2006turb}. As the ion Larmor radius 
is many orders of magnitude smaller than the collisional mean free path, 
the plasma is weakly collisional, which results in conservation of adiabatic 
invariants, the first of them being the magnetic moment of a particle 
$\mu = v_{\perp}^2 / (2 B)$, where $v_{\perp}$ is the component of the particle 
velocity perpendicular to the magnetic field. Consequently, the magnetic-field 
strength changes are correlated with changes in the perpendicular pressure, 
giving rise to pressure anisotropy. In turn, pressure anisotropy triggers 
firehose and mirror instabilities \citep[][]{Chandrasekar1958,Parker1958,
Hasegawa} that hold the degree of anisotropy $\Delta=(p_{\perp} - p_{\parallel})
/p_{\perp}$ at marginal levels $|\Delta| \sim 1/\beta$, where $\beta$ is the 
plasma beta, the ratio of thermal to magnetic energy density (for observational 
evidence in the solar wind, see 
\citealt{Kasper2002,Hellinger2006,Bale2009}; for theoretical discussion of 
possible mechanisms of maintaining marginality, see \citealt{Melville2015} and 
references therein). The firehose instability occurs when $\Delta < -2/\beta$, 
which happens in regions where the field strength is decreasing, near the 
reversal points of the field lines, and typically generates small 
($\delta B_{\perp} / B \ll 1$) transverse Alfv$\acute{\rmn{e}}$nic fluctuations of 
the field direction. The mirror instability (or the `mirror mode') is a 
resonant instability set off when $\Delta > 1/\beta$, which is the case 
where the field is amplified along the stretches of the field lines. The 
mirror mode produces fluctuations of magnetic-field strength of order unity 
($\delta B / B \sim 1$), which form magnetic traps and may, 
in principle, inhibit electron transport along the field lines. While 
field-strength fluctuations $\delta B / B \sim 1$ can 
also be generated by turbulent motions, we will argue in Section~\ref{sec:mhd} 
that the resulting suppression of transport is very moderate, because 
the electron mean free path $\lambda$ is smaller than the parallel 
correlation length of the magnetic field $l_B$, and the electrons can 
escape from magnetic traps relatively easily. Illustratively, the 
presumed combined spectrum of magnetic-field strength fluctuations in the ICM is 
sketched in Fig.~\ref{fig:Bpar_sp}: the magnetic mirrors capable of 
efficient suppression of electron transport reside in the region 
$\lambda \gg l_B$.  

The mirror magnetic fluctuations are impossible to observe directly 
in the ICM due to their extremely small scales, but they can be modeled 
by numerical simulations. The recent hybrid particle-in-cell simulations of the 
firehose and mirror instabilities in a shearing box done by \cite{Kunz2014} 
suit this task well in providing the typical statistical 
properties of the magnetic mirror fluctuations. In this paper, we use 
the mirror fluctuations produced by their simulations to model the 
electron motion along the resulting perturbed field lines and 
estimate the conductivity.

The paper is organized as follows. In Section~\ref{sec:model}, we 
describe a model for parallel electron diffusion and its Monte Carlo 
equivalent for numerical calculations. Then, in \Secref{sec:mirror}, 
we apply this model to the mirror magnetic fluctuations taken from 
the simulations of \cite{Kunz2014} to infer the suppression of 
parallel electron diffusivity (\Secref{sec:mirror_results}) and thermal 
conductivity (\Secref{sec:cond}). Next, in \Secref{sec:mhd}, we argue 
that large-scale turbulent magnetic fluctuations in the ICM, 
modeled by an isotropic MHD simulation, do not cause a sizable 
suppression. Finally, in \Secref{sec:disc}, we summarize our results 
and their relevance to the problem of thermal conduction in the ICM 
and in turbulent weakly collisional plasmas in general.

\section{A model for parallel electron diffusion in a static magnetic field}
\label{sec:model}

For our calculations, we assume the electron diffusion timescale 
to be smaller than the characteristic times of fluid motions and 
of the magnetic-field evolution, so that the magnetic field can be viewed 
as static. We will assess the validity of this assumption in 
Section~\ref{sec:disc}. 

If magnetic fluctuations occur at parallel scales $l_B$ much larger than the 
electron Larmor radius $\rho_e$, which is indeed true for turbulent 
magnetic fluctuations, as well as for the mirror-mode perturbations 
produced at the scale of $1-100~\rho_i$, where $\rho_i = (m_i/m_e)^{1/2} 
\rho_e \sim 40~\rho_e$ is the ion Larmor radius, and if all fluid motions 
are neglected, we may use the drift-kinetic equation
\beq
\label{eq:GK}
\frac{\partial f}{\partial t} + v \xi \nabla_{\parallel} f - \frac{\nabla_{\parallel} B}{B} v \frac{1-\xi^2}{2} 
\frac{\partial f}{\partial \xi} = \nu (v) \frac{\partial}{\partial \xi} \frac{1-\xi ^2}{2} 
\frac{\partial f}{\partial \xi}, 
\eeq
to evolve the electron distribution function $f=f(t,\vc{x},v,\xi)$ 
\citep{Kulsrud1964}. Here $\nabla_{\parallel}=\hat{\vc{b}} \cdot \vc{\nabla}$ 
is the derivative taken along the local magnetic field and $\xi=\hat{\vc{b}} 
\cdot \vc{v} / v = \cos\theta$, where $\theta$ is the pitch angle. The unit 
vector $\hat{\vc{b}} = \vc{B}/B$ points in the local magnetic-field direction. 
The last term on the left-hand side of equation~(\ref{eq:GK}) represents the 
mirror force, which guarantees conservation of the magnetic moment $\mu =  
v_{\perp}^2 / (2B) = v^2 (1-\xi^2)/(2B)$ in the absence of collisions. 
Isotropic collisions with collision frequency $\nu$ are described by the 
Lorenz pitch-angle scattering operator on the right-hand side of equation~
(\ref{eq:GK}). In this section, we restrict our analysis to monoenergetic 
electrons, so there is no energy exchange between the particles. We also 
neglect the electric field because, close to marginal stability ($\Delta 
\sim 1/\beta$), the mirror instability generates an electric field of 
order $E_{\parallel} \sim (T/e) (\nabla_{\parallel} B/B)(1/\beta)$, where 
$T$ is the electron temperature, $e$ the absolute electron charge. In 
astrophysical plasmas, $\beta$ is typically large (e.g., $\sim 100$ in the 
ICM), so the electric field can be safely neglected. 

The problem is effectively one-dimensional with respect to the arc 
length $\ell$ along a field line, because all spatial derivatives in 
equation (\ref{eq:GK}) are taken along the local magnetic field. Thus, 
we can rewrite equation~(\ref{eq:GK}) in field-aligned coordinates by 
normalizing the distribution function using the Jacobian of this 
coordinate transformation, $\widetilde{f}(t,\ell,\xi) 
= f/B$:
\beq
\label{eq:GK2}
\frac{\partial \widetilde{f}}{\partial t} + \xi \frac{\partial \widetilde{f}}{\partial\ell} - 
M(\ell) \frac{\partial}{\partial \xi} \frac{1-\xi^2}{2} \widetilde{f} = 
\frac{1}{\lambda} \frac{\partial}{\partial \xi} \frac{1-\xi ^2}{2} \frac{\partial \widetilde{f}}{\partial \xi}, 
\eeq
where $M(\ell)=\partial\ln{B} / \partial\ell$ is the mirror force, 
$\lambda=v/\nu$ is the electron mean free path\footnote{Electron 
collisionality may be anomalous due to, e.g., scattering off magnetic 
fluctuations generated by electron microinstabilities. This could reduce 
the effective electron mean free path, but our main results would remain 
valid as long as the effective mean free path is much larger than 
the ion Larmor scale.}, and time has been rescaled as $vt \rightarrow t$. Using 
$\xi = \cos{\theta}$, the distribution function $F(t,\ell,\theta) = 
\widetilde{f} \sin{\theta} $ satisfies
\beq
\frac{\partial F}{\partial t} + \cos{\theta} \frac{\partial F}{\partial \ell} + \frac{\partial}{\partial \theta} 
\left [ \frac{1}{2} M(\ell) \sin{\theta} + \frac{\cot{\theta}}{2 \lambda} \right ] F = \frac{1}{2 \lambda} 
\frac{\partial ^2 F}{\partial \theta ^2}.
\eeq
A convenient way to solve this equation by the Monte Carlo method is to 
treat it as the Fokker--Planck equation for particles whose equations of 
motions are
\bea
\nonumber
\label{eq:part}
\dot{\ell} &=& \cos{\theta},\\
\dot{\theta} &=& \frac{1}{2} M(\ell) \sin{\theta} + \frac{\cot{\theta}}{2 \lambda} + 
\frac{1}{\sqrt{\lambda}} \eta (t),
\eea
where $\eta (t)$ is a unit Gaussian white noise, $\langle \eta(t) 
\eta(t') \rangle = \delta(t-t')$. As clearly seen from these equations, 
a particle experiences the mirror force $M(\ell)$ defined by the static 
magnetic field and isotropizing collisions represented by the last two 
terms on the right-hand side. Equations~(\ref{eq:part}) can be easily 
solved numerically. 

Without collisions, only the particles in the loss cone defined by 
$|\xi| > (1-2 \mu B /v^2)^{1/2}$ can travel freely. The rest are 
reflected by regions of strong field (magnetic mirrors). Collisions 
allow trapped particles to get scattered into the loss cone and 
escape from magnetic traps. Oppositely, a free particle can be 
knocked out of the loss cone by collisions and become trapped. The 
key parameter that defines the regime of diffusion is the ratio of 
the collisional mean free path $\lambda$ to the parallel correlation 
length of the magnetic field $l_B$. If $\lambda/l_B \ll 1$, collisions 
make magnetic trapping ineffective, and the electrons undergo ordinary 
diffusion with diffusion coefficient $D \sim \lambda v$. In the opposite 
limit $\lambda / l_B \gg 1$, collisions are very rare, so the pitch 
angle changes only slightly over the correlation length of the field. 
In this regime, the suppression of diffusion is greatest because a 
certain fraction of the particles is trapped and, in addition, the 
passing particles have their mean free paths effectively reduced as 
small-angle collisions cause leakage from the loss cone so that a 
free particle travels only a fraction of its mean free path before 
it is scattered out of the loss cone and becomes trapped \citep{CC1998,
Chandran1999}.

\section{Electron diffusion in a magnetic mirror field}
\label{sec:mirror}

\subsection{Properties of the mirror field}

\begin{figure}
\centering
\includegraphics[width=84mm]{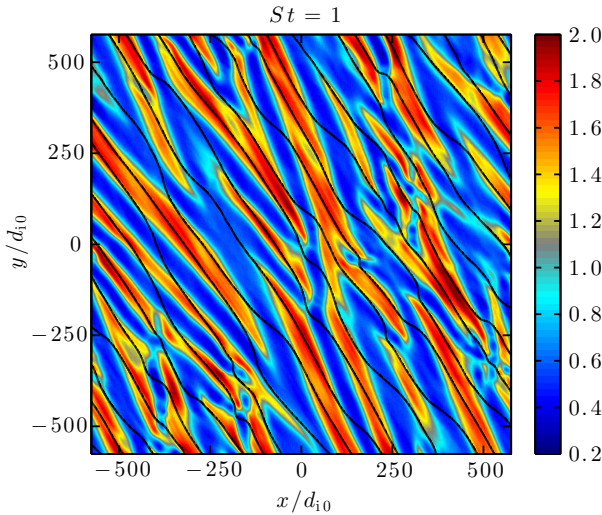}
\caption{Spatial structure of the mirror instability \citep{Kunz2014} 
during the secular phase of the perturbations' growth, after one shear 
time. The magnetic-field strength $B/\langle B \rangle$ is shown by color, 
the field lines are shown by contours. Length is in the 
units of the ion skin depth $d_{i0} = \rho_{i}(St=0) / \sqrt{\beta_0}$. 
At time $St=1$, the ion Larmor radius is $\rho_i\approx 8.7 d_{i0}$.}
\label{fig:mirror_struct}
\end{figure}

A description of the numerical code and set-up used to generate the 
mirror magnetic fluctuations can be found in \cite{Kunz2014}. The code 
(\citealt{Kunz2014a}a) is a hybrid-kinetic particle-in-cell code, in which the 
electrons are fluid while the ions are treated kinetically as quasi-particles. 
To trigger the mirror instability, a square 2D region of plasma of spatial 
extent $L =1152d_{i0}$, where $d_{i0}$ is the initial ion skin depth, is 
threaded by a magnetic field directed at an angle to the $y$-direction 
and subjected to a linear shear $\vc{u}_0 = - S x \vc{\hat{y}}$, 
which stretches the field lines and, by adiabatic invariance, produces 
pressure anisotropy. The initial magnetic field strength is $B_0$, the 
initial plasma beta of the ions is taken 
to be $\beta_0=200$, and the shear is $S=3\times10^{-4} \Omega_i$, where $\Omega_i$ is 
the ion gyrofrequency. The ion Larmor radius is $\rho_i=\sqrt{\beta} d_i$. 
Once the (ion) pressure anisotropy $\Delta=p_{\perp}/p_{\parallel}-1$ 
reaches $1/\beta$, the plasma becomes mirror-unstable. Magnetic perturbations 
grow exponentially until they become large enough to drive the anisotropy 
back to the marginal level, $\Delta \rightarrow 1/\beta$. Persistent 
large-scale driving of the pressure anisotropy, coupled with the requirement
for the plasma to remain marginally stable, leads to a long phase of 
secular growth of the mirror perturbations. The spatial structure of the perturbations during this phase 
is shown in Fig.~\ref{fig:mirror_struct}. 
The mirror 
fluctuations are elongated in the direction of the mean magnetic 
field and have $\delta B_{\parallel} \gg \delta B_{\perp}$. During 
this secular phase, the field grows as $\delta B \propto 
t^{4/3}$ and the dominant modes shift towards longer wavelengths 
($k_{\parallel} \rho_i \sim 10^{-2}$) as the pressure anisotropy asymptotically 
approaches marginal stability. The marginal stability is achieved and 
maintained during the secular phase by the trapping of ions in magnetic 
mirrors \citep[see][]{Rincon2015,Melville2015}. The final saturation sets 
in when $\delta B/B_0 \sim 1$ at $St \gtrsim 1$, and is caused by the 
enhanced scattering of ions off sharp ($\delta B_{\parallel}/B_0 \sim 1$, 
$k_{\parallel} \rho_i \sim 1$) bends in the magnetic field at the edges 
of the mirrors. 

\begin{figure}
\begin{center}
\includegraphics[width=76mm]{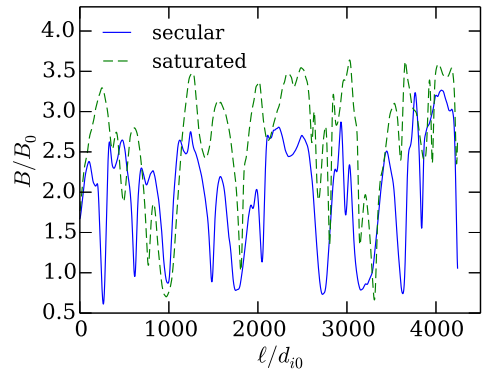}
\caption{Variation of the magnetic-field strength along a field 
line during the two different phases of evolution of the mirror 
fluctuations ({\it solid}: secularly growing fluctuations 
at time $St=1$; {\it dashed}: saturated fluctuations 
at time $St\approx 1.8$). $B_0$ is the initial 
magnetic-field strength in the simulation. For convenience, 
4000-$d_{i0}$ line segments are shown. The ion Larmor radii are 
$\rho_i(St=1) \approx 8.7d_{i0}$ and $\rho_i(St \approx 1.8) \approx 6.2 d_{i0}$.}
\label{fig:field_lines}
\end{center}
\end{figure}

\begin{figure}
\begin{center}
\includegraphics[width=76mm]{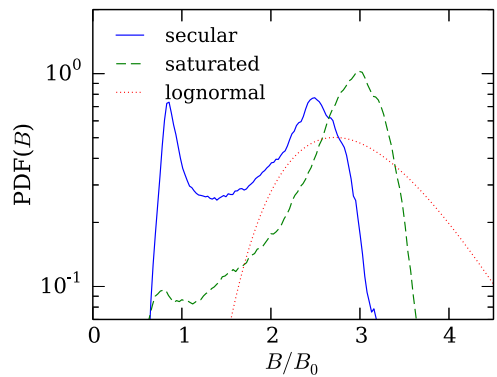}
\caption{PDFs of the magnetic-field-strength fluctuations generated by 
the mirror instability. A lognormal distribution with rms equal 
to the rms of the logarithm of the saturated mirror fluctuations 
is shown by the dotted line for comparison.}
\label{fig:mirror_pdf}
\end{center}
\end{figure}

We note that the electrons in the code are isothermal with $T_e=T_i$, 
so we are not attempting to solve the problem of the electron heat 
transfer self-consistently (no thermal gradients and heat fluxes are 
present). We have extracted two representative magnetic-field lines 
from the simulation domain, one during the secular phase ($St=1$), 
and one during the saturated phase ($St \approx 1.8$). Each of these
crosses the box eight times (note that, although the box is shearing-periodic, 
a field line does not bite its tail and hence can be followed over 
several crossings) and has a length of $\approx 18000d_{i0}$ (we 
adopt $d_{i0}$ as our unit of length because $d_i$ is practically 
constant in time, while $\rho_i$ is a function of the field strength). 
The variation of the magnetic-field strength $B$ along the lines is 
shown in Fig.~\ref{fig:field_lines}. From the analysis of the probability 
density functions (PDF) of $B$ for the two field lines (Fig.~
\ref{fig:mirror_pdf}), it is clear that in both cases, the PDFs have 
abrupt cut-offs at large $B \sim$ several $B_0$, as well as at small 
$B$. Therefore, the field is bounded with no large deviations from 
the mean value, in contrast to, e.g., a lognormal stochastic magnetic 
field with the same rms (shown by the dotted line in Fig.~\ref{fig:mirror_pdf} 
for comparison) with a tail in its PDF, for which there is always a 
non-zero probability to find a large deviation at a large enough scale. 
This also means that the extracted field lines fully represent the 
statistics of $B$ (a longer field line would not contain more statistical 
information). The bounded PDF($B$) is a key property of the mirror 
magnetic field, which leads to a finite value of suppression of 
electron transport at large $\lambda/l_B$, unlike in the case of 
stochastic magnetic fields (lognormal, Gaussian, exponential) that can 
completely inhibit particle transport in this limit \citep{MalyshkinKulsrud2001, 
Albright2001}. 


\subsection{Suppression of electron diffusivity in the limit $\lambda 
/ l_B \gg 1$}
\label{sec:mirror_results}
\subsubsection{Results of the Monte Carlo simulations}
\label{sec:MCres}
For the two extracted field lines, we integrate the particles' 
trajectories defined by equations~(\ref{eq:part}) numerically. 
Initially, the particle distribution is isotropic with the particle 
density along a field line set to $\propto 1/B$, which is a uniform 
density distribution in real space [recall the Jacobian of the 
coordinate transformation to field-aligned coordinates in 
equation~(\ref{eq:GK2})]. Then we trace the evolution of the 
particles over time $t_1=20~t_{\rmn{coll}}$, where $t_{\rmn{coll}}
=1/\nu$ is the collision time. The monoenergetic diffusion 
coefficient $D$ is calculated as 
\beq
\label{eq:DMC}
D=\frac{\langle [\ell_i(t_1)-\ell_i(t_0) ]^2 \rangle }{2 (t_1-t_0) },
\eeq
where $\ell_i$ are the particles' displacements. We choose $t_0=5~
t_{\rmn{coll}}$ in order to allow the particles to collide a few 
times until the ballistic regime gives way to diffusion at $t \gtrsim 
t_{\rmn{coll}}$. The same procedure is carried out for several different 
ratios $\lambda / l_B$ in the range $2 \times 10^{-4}$--$5 \times 10^4$. 
The correlation lengths of the field strength along the lines are $l_B
\approx 850d_{i0}\approx 100\rho_i(St=1)$ for the secular phase and 
$l_B\approx 1430d_{i0}\approx 230\rho_i(St \approx 1.8)$ for the saturated phase. 

\begin{figure}
\begin{center}
\includegraphics[width=76mm]{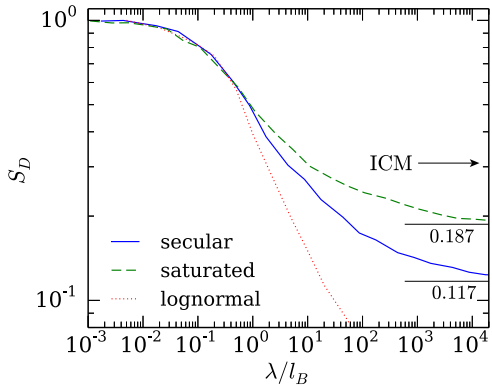}
\caption{Suppression factor of the electron diffusivity $S_D=D/D_0$ 
in the secularly growing (solid line) and saturated (dashed line) 
magnetic mirror fluctuations. The correlation lengths of the 
field strength along the field lines are $l_B \approx 850d_{i0}\approx 
100\rho_i(St=1)$ during the secular phase and $l_B\approx 1430d_{i0} \approx 
230\rho_i(St \approx 1.8)$ for the saturated mirrors. For comparison, the dotted line shows 
suppression in a synthetic lognormal magnetic field with the same rms value 
of $\log{B}$ as during the phase of secular growth of the mirror 
fluctuations, correlation length $l_B=850d_{i0}$ and a Kolmogorov 
spectrum in space (see its PDF in Fig.~\ref{fig:mirror_pdf}).}
\label{fig:suppr}
\end{center}
\end{figure}

Defining $D_0=(1/3) \lambda v$, the diffusion coefficient in the absence 
of the magnetic fields, we thus obtain the monoenergetic diffusion 
suppression factor $S_D=D/D_0$ as a function of $\lambda / l_B$ (Fig.~
\ref{fig:suppr}). Averaging the monoenergetic diffusivity $D$ over a 
thermal distribution of the electron speeds $v$ introduces only a 
slight change in the shape of the function $S_D(\lambda/l_B)$, so 
we only present the monoenergetic diffusion suppression in what 
follows.
 
For magnetic mirror fluctuations in the ICM, the limit $\lambda 
/ l_B \gg 1$ is the relevant one, because the ion Larmor radius 
is many orders of magnitude smaller than the mean free path:
\bea
\lambda &\approx & 20~\mathrm{kpc} \left ( \frac{T}{8~\mathrm{KeV}} \right )^2 
										 \left (\frac{n}{10^{-3}~\mathrm{cm}^{-3}} \right )^{-1},\\
\rho_i &\approx & 5\times 10^{-12}~\mathrm{kpc} \left ( \frac{T}{8~\mathrm{KeV}} \right )^{1/2} 
											 \left ( \frac{B}{1~\mu \mathrm{G}} \right ) ^{-1}.
\eea
In this limit, the suppression factor asymptotically approaches 
$S_D \approx 0.12$ during the secular phase, and $S_D\approx0.19$ 
for the saturated mirrors. The absence of the mean-free-path dependence 
at large $\lambda/l_B$ is due to the fact that the mirror 
fluctuations are bounded, as we have seen by analyzing their 
PDF (Fig.~\ref{fig:mirror_pdf}). This is very different from the 
case of stochastic magnetic mirrors (see \citealt{MalyshkinKulsrud2001}; 
we will discuss stochastic magnetic mirrors in more detail in 
Section~\ref{sec:stochastic}). 

\subsubsection{The role of the PDF($B$)}
\label{sec:PDF}

\begin{figure}
\begin{center}
\includegraphics[width=76mm]{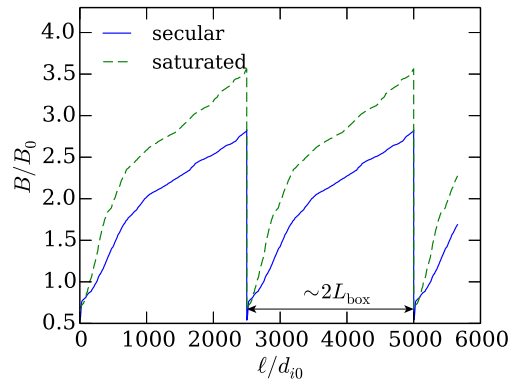}
\caption{Equivalent representation of the mirror magnetic fluctuations 
along the field lines in the case $\lambda/l_B \gg 1$. Reordering the 
magnetic-field strength values over length $L\sim 2L_{\rmn{box}}$, 
$l_B \lesssim L \ll \lambda$, does not change the PDF($B$) and, therefore, 
the amount of diffusion suppression.}
\label{fig:equiv}
\end{center}
\end{figure}

The limiting values of $S_D$ in Fig.~\ref{fig:suppr}
depend only on the PDF($B)$ along the field lines. 
This fact is intuitive because the change in the pitch angle of 
a passing particle due to collisions as it travels the correlation 
length $l_B$ of the field is very small, and the order in which the 
particle encounters regions of different $B$ plays no role. Therefore, 
one can rearrange the mirror magnetic fluctuations (Fig.~\ref{fig:field_lines}) 
by sorting the array elements in ascending order over some length $L$, 
$l_B \lesssim L \ll \lambda$, and making the resulting array periodic 
with period $L$ (Fig.~\ref{fig:equiv}). Since the field is bounded, 
we do not lose statistical information if $L$ is set to just a few $l_B$. 
Clearly, this procedure keeps the PDF($B$) unchanged, and the 
resulting magnetic field produces the same amount of suppression, 
while having a much simpler shape. By comparing such simplified shapes 
of magnetic fluctuations for different field lines, one can determine 
which line causes more suppression. The loss cone for a particle is 
defined as $|\xi| > (1-B/B_{\rmn{max}})^{1/2}$, where $B$ is the 
magnetic-field strength at the location of the particle and $B_{\rmn{max}}$ 
is the global maximum of the field strength. The more concave (or less convex in 
our case of the mirror fluctuations) the shape is, the more suppression 
is produced by magnetic mirroring, because the loss cones of most of the 
particles become narrower. The extreme case of this is a field that consists 
of narrow periodic peaks of height $B_{\rmn{max}}-B_{\rmn{min}}$. This 
field causes maximum suppression of diffusion for given values of 
$B_{\rmn{min}}$ and $B_{\rmn{max}}$, because the loss cone for almost 
all the particles, $|\xi| > (1-B_{\rmn{min}}/B_{\rmn{max}})^{1/2}$, 
is the narrowest it can be for all possible PDF($B$). Based on this argument, 
and noticing that, in Fig.~\ref{fig:mirror_pdf}, the PDF of the saturated 
mirror field is more concentrated around the maximum field strength, one can 
predict more suppression of electron diffusion by secularly growing mirrors 
than by saturated ones, even though $B_{\rmn{max}}$ is smaller for the former 
than for the latter. Indeed, we see in Fig.~\ref{fig:suppr} that the effect 
of the field shape prevails over the difference in $B_{\rmn{max}}$, and the 
diffusion suppression is stronger for the secular phase.

\subsubsection{The physical mechanism of the suppression of electron diffusion 
at large $\lambda / l_B$} 
\label{sec:two_eff}

In the limit $\lambda / l_B \gg 1$, \cite{Chandran1999} derived an 
analytic expression for the suppression of diffusivity of monoenergetic 
electrons by periodic magnetic mirrors [their equation~(95)]:
\beq
\label{eq:RDCC}
S_D = \frac{3}{\langle 1/B' \rangle} \int_0^1 d \mu'_1 \int_{\mu'_1}^
1 d \mu'_2 
\frac{1}{\langle |\xi (\mu'_2)| /B' \rangle},
\eeq
where $B'=B/B_{\rm max}$ is the magnetic-field strength normalized to its 
global maximum value,  $\mu'=\mu/\mu_{\rmn{crit}}$ is the magnetic moment of a particle 
$\mu=v^2 (1-\xi^2) / (2B)$ normalized to $\mu_{\rm crit} = v^2 / (2B_{\rm max})$, 
the averaging is performed over the period of the magnetic field, and the 
integration is carried over the passing particles in the loss cone. 
As we have discussed in Section~\ref{sec:PDF}, 
bounded mirror fluctuations can be replaced by periodic variations with the 
same PDF($B)$ without affecting the suppression factor. This means that 
equation~(\ref{eq:RDCC}), where the averaging in the angle brackets is done 
over PDF($B)$, can be readily applied to the simulated mirror 
fluctuations. The asymptotic values of $S_D$ calculated by \eqref{eq:RDCC}, 
$S_D\approx 0.117$ for the secularly growing mirrors and $S_D\approx 0.187$ for the 
saturated ones, agree extremely well with the results of our Monte Carlo simulation 
(see Fig.~\ref{fig:suppr}). 

We can break down the suppression effect encoded in equation~(\ref{eq:RDCC}) into 
two physical effects: 
\beq
\label{eq:suppr_two_eff}
S_D = S_p \frac{\lambdaeff}{\lambda},  
\eeq
where $S_p$ is the suppression of diffusivity due to the fact that 
only the passing particles contribute to electron transport, $\lambdaeff$ 
is the effective mean free path of the passing particles, reduced 
because a passing particle is scattered out of the loss cone, becomes 
trapped and randomizes its direction of motion in only a fraction 
of its collision time. The parameters $S_p$ and $\lambdaeff$ have a 
very clear physical interpretation in terms of the particle velocity 
autocorrelation function.

\begin{figure}
\begin{center}
\includegraphics[width=80mm]{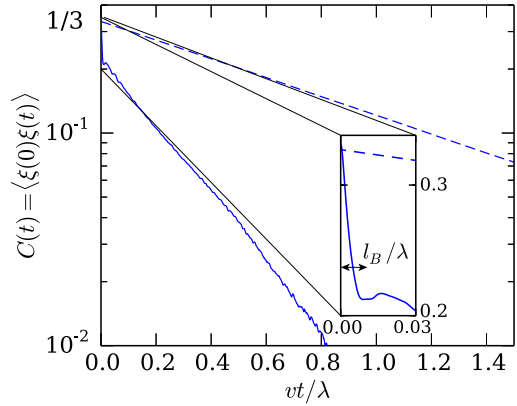}
\caption{The parallel velocity autocorrelation functions $C(t)/v^2$ 
for electron diffusion in a constant magnetic field (dashed) and in 
the magnetic field of secularly growing mirrors (solid), based on the results 
of the Monte Carlo simulation. The ratio of the electron mean free 
path to the correlation length of the magnetic field $\lambda/l_B=100$. 
The narrow peak at $t<l_B/v$ is caused by the bouncing trapped particles, which do not 
contribute to electron transport. } 
\label{fig:ac}
\end{center}
\end{figure}

The electron diffusion coefficient $D$ can be expressed as 
the integral of the parallel-velocity 
autocorrelation function $C(t)$:
\beq
\label{eq:DCt}
D = \int_0^{\infty} \langle v_{\parallel}(0) v_{\parallel}(t) \rangle dt 
		\equiv \int_0^{\infty} C(t) dt.
\eeq 
Using the results of our Monte Carlo simulations of monoenergetic 
diffusion in magnetic fluctuations generated by the mirror instability, 
we can calculate the parallel-velocity autocorrelation function, which 
is, for monoenergetic electrons, the autocorrelation function of the 
cosine of the pitch angle $\xi=\cos{\theta}$, namely 
$C(t) = v^2 \langle \xi(0) \xi(t) \rangle$. It is plotted in Fig.~
\ref{fig:ac} for diffusion with no magnetic mirrors (dashed line) and 
diffusion in the mirror magnetic field with $\lambda / l_B = 100$ 
(solid line). With no mirrors, 
\beq
C_0(t)= \frac{1}{3} v^2 e^{-\nu t},
\eeq
where $\nu = 
v/\lambda$. Both with or without mirrors, the autocorrelation function 
is equal to $1/3$ at $t=0$ due to isotropy (even in the presence of 
magnetic mirrors, collisions restore isotropy over times $\gg 
\nu^{-1}$). In the mirror field, $C(t)$ has a narrow peak of width 
$\sim l_B/v$ at small $t$, while the rest of the autocorrelation 
function is an exponential that is well fitted by 
\beq
\label{eq:Ct}
C(t>l_B/v) = \frac{1}{3} S_p v^2 e^{-\nu_{\rmn{eff}} t},
\eeq
where $S_p<1$ and $\nu_{\rm eff} > \nu$.

The narrow peak of $C(t)$ at small $t$ is caused by the contribution 
to $C(t)$ of the population of trapped particles, which bounce inside 
magnetic traps at a typical time scale $\sim l_B/v$. The physical meaning 
of the reduction factor $S_p<1$ is that only the passing articles contribute to transport 
processes. This factor can be calculated as
\beq
\label{eq:Sp}
S_p = f_{\rmn{pass}} \frac{\langle \xi^2 \rangle_{\rmn{pass}}}{\langle \xi^2 \rangle}, 
\eeq
where $f_{\rmn{pass}}$ is the fraction of the passing particles, 
the averaging is performed over the passing particles in the 
numerator, and over all particles in the denominator. The value 
of $S_p$ is greater than simply the fraction of the passing 
particles, because they travel in their loss cones and, therefore, 
have parallel velocities greater than the mean square parallel 
velocity $v^2{\langle \xi^2 \rangle} = v^2/3$ averaged over all particles.

The physical interpretation of the fact that $\nu_{\rmn{eff}} > 
\nu$ in equation~(\ref{eq:Ct}) is the reduced effective mean free 
path of the passing particles: recall that a passing particle travels 
only a fraction of $\lambda$ before it becomes trapped and, 
obviously, $\nu_{\rmn{eff}}/\nu = \lambda/\lambdaeff$. Since the diffusion 
coefficient is the integral of $C(t)$ [equation~(\ref{eq:DCt})], 
we obtain equation~(\ref{eq:suppr_two_eff}).

From the above arguments, it follows that a system with magnetic mirrors 
and $\lambda/l_B \gg 1$ can be translated into a system with no mirrors, 
but with a lower mean square parallel velocity and an enhanced scattering 
rate. The lower parallel velocity is related to the fact that the passing 
particles become trapped now and then, and while they are trapped, their 
effective parallel velocity is zero. 

Using the results of our Monte Carlo simulations and the velocity 
autocorrelation function analysis described above, we can measure $S_p$ and 
$\lambda_{\rm eff}/\lambda$. For the secularly growing mirrors, we get 
$S_p\approx0.63$, $\lambda_{\rm eff}/\lambda \approx 0.19$; for the 
saturated mirrors, $S_p \approx 0.74$, $\lambda_{\rm eff}/\lambda 
\approx 0.26$. Substituting these into equation~(\ref{eq:suppr_two_eff}), 
we recover $S_D\approx0.12$ and $S_D\approx0.19$, the same as was 
obtained in direct measurement [equation~(\ref{eq:DMC})] and from 
equation~(\ref{eq:RDCC}).


\subsection{Suppression of electron thermal conductivity in the limit $\lambda / l_B \gg 1$}
\label{sec:cond}

As we have shown above, the effect of magnetic mirrors on scales much 
larger than the electron mean free path is the suppression of spatial 
diffusion via two effects: reduced fraction and reduced effective mean free 
path (or, equivalently, an enhanced scattering rate) 
of the passing particles participating in transport 
[see \eqref{eq:suppr_two_eff}]. In a plasma 
with a temperature gradient and no mirrors, heat transport is governed not 
only by pitch-angle diffusion, but by diffusion in the energy space as well. 
Magnetic mirrors do not change a particle's energy, therefore, one can 
model their effect on large scales by enhancing the pitch-angle scattering 
rate (but not the energy diffusion rate) and, simultaneously, reducing the 
effective density of particles 
carrying energy in order to subtract the trapped population. 

The rates of pitch-angle scattering (perpendicular velocity diffusion) 
$\nu_{\perp}$ and energy exchange $\nu_{\varepsilon}$ for a test electron in 
a hydrogen plasma are \citep{Spitzer1962}
\bea
\nu_{\perp, es} &=& 2 [(1-1/2x) \psi(x) + \psi'(x)] \nu_0,\\
\nu_{\varepsilon, es} &=& 2 [(m_e/m_s) \psi(x) - \psi'(x)] \nu_0,
\eea
where
\bea
\lefteqn{\nu_0 = \frac{4 \pi e^4 \ln{\Lambda} n_e}{m_e^2 v^2}, \hspace{5mm} 
		x = v^2/v_{{\rm th},e}^2,}&&\\
\lefteqn{\psi(x) = \frac{2}{\sqrt{\pi}} \int_0^x dt \sqrt{t} e^{-t}, \hspace{5mm} 
		\psi'(x) = \frac{d\psi}{dx},}&&
\eea
$s=e,i$ is the species of the background particles, $v_{{\rm th},e} = 
(2kT_e/m_e)^{1/2}$ is the electron thermal speed, and $\ln{\Lambda} 
\sim 40$ is the Coulomb logarithm. Heat is transferred by slightly superthermal 
electrons with $v\approx 2.5~v_{{\rm th},e}$ (this rough estimate is based 
on a simple calculation of thermal conductivity for a Lorenz gas, 
when electrons interact only with ions). At this velocity, $\nu_{\perp,ei}
\approx 1.5~\nu_0$, $\nu_{\perp, ee} \approx 1.8~\nu_0$, $\nu_{\varepsilon}= 
\nu_{\varepsilon, ei} + \nu_{\varepsilon, ee} \approx \nu_{\varepsilon, ee} 
\approx 2~\nu_0$. Thus, the electron energy exchange rate $\nu_{\varepsilon}$ 
is close to the total perpendicular electron diffusion rate $\nu_{\perp} 
\approx 3.3~\nu_0$ for the heat-conducting electrons. 


At this point, we make a qualitative assumption that the total rate 
of spatial energy transfer can be reasonably approximated by the sum 
of the energy exchange rate and the pitch-angle scattering rate. 
This assumption is corroborated by the mathematical fact that in a plasma 
with a gradient of a diffusing passive scalar, the flux of the scalar 
is inversely proportional to the sum of the rate of spatial diffusion of 
the particles (or pitch-angle scattering) and the collisional exchange rate 
of the passive scalar [see Appendix~\ref{app:A}, \eqref{eq:kappaa0}]. 
The passive scalar in this calculations models temperature, as if 
every particle carried an averaged value of thermal energy that 
did not depend on the particle's velocity. 
Then for the thermal conductivity $\kappa$, we use the approximation
\beq
\kappa \sim \frac{ v_{{\rm th},e}^2}{\nu_{\varepsilon}+\nu_{\perp}}.
\eeq  

The reduction of the effective density of the heat-conducting 
electrons affects both pitch-angle and energy diffusion, while 
the enhanced scattering rate only affects pitch-angle diffusion. 
Thus, the suppression of thermal conduction is smaller than the 
suppression of spatial diffusion. A qualitative expression for 
the suppression of thermal conductivity $S_{\kappa} = \kappa/\kappa_0$ 
in the limit $\lambda / l_B \gg 1$ is then
\beq
\label{eq:diffcond}
S_{\kappa} \sim S_p \frac{\nu_{\perp} + \nu_{\varepsilon}}{(\lambda/\lambdaeff) 
			\nu_{\perp} + \nu_{\varepsilon}} \sim 
				\frac{2S_p}{1 + \lambda/\lambdaeff}, 
\eeq
where $S_p$ and $\lambdaeff$ are the parameters in \eqref{eq:suppr_two_eff}: 
$S_p$ is related to the fraction of the passing particles, $\lambdaeff$ is the 
effective mean free path of the passing particles. For a passive scalar in the 
limit $\lambda/l_B\gg1$, \eqref{eq:diffcond} is exact and derived in 
Appendix~\ref{app:A} by establishing a simple relationship between the amount of 
suppression of the scalar flux and the parallel-velocity autocorrelation function. 
Substituting the values of $S_p$ and $\lambdaeff/\lambda$ calculated at the end of 
Section~\ref{sec:two_eff} into \eqref{eq:diffcond}, we obtain the suppression factor of 
thermal conductivity: $S_{\kappa} \sim 0.2$ for the secularly growing mirrors 
and $S_{\kappa} \sim 0.3$ for the saturated ones. We see that heat transport is 
suppressed by a factor of $\sim 2$ less than spatial diffusion, because the diffusion 
in energy space is suppressed much less than the spatial (pitch-angle) diffusion. 
Equation~(\ref{eq:diffcond}) can only be used when $S_p$ and $\lambdaeff$ 
do not depend on the electron velocity (or, equivalently, on the electron mean 
free path $\lambda$), which is indeed the case in the limit $\lambda / l_B \gg 1$ (see Fig.
~\ref{fig:suppr}).

\section{Electron transport in MHD turbulence}
\label{sec:mhd}
As mentioned in Section~\ref{sec:intro}, another source of
fluctuations of magnetic-field strength in the ICM is turbulent 
stretching/compression of the field lines. The turbulent dynamo produces 
a stochastic distribution of the field strength along a field line: 
lognormal during the kinematic phase, exponential in saturation \citep[see][]
{SchekCowleyTaylor2004}. However, as we have also noted in Section~\ref{sec:intro}, 
we do not expect much suppression of thermal conduction by these fields, 
because the electron mean free path is smaller than the parallel correlation 
length of turbulent magnetic fluctuations, and so magnetic mirrors are rare and not 
very effective. In this section, we demonstrate this explicitly by means of an 
isotropic MHD simulation of turbulent dynamo.

\subsection{Suppression of electron transport in a system of stochastic magnetic mirrors} 
\label{sec:stochastic}
Before we consider the magnetic fields produced by the turbulent MHD dynamo, 
let us first illustrate how different diffusion in a stochastic magnetic 
field is from the case of a periodic field (characteristic of the mirror 
fluctuations) by the example of a lognormally distributed field. A 
stochastic magnetic field with a long tail in its PDF produces a 
larger amount of suppression compared to periodic magnetic fluctuations, 
because the dominant suppression is caused by the so-called 'principal 
magnetic mirrors' of strength $m_p=B_p /\langle B \rangle \gg 1$ separated 
from each other by a distance of order the effective mean free path 
(characteristic length that a passing particle travels before it gets 
scattered out of the loss cone and becomes trapped) $\lambda/m_p$ 
\citep{MalyshkinKulsrud2001}. Because $\lambda/m_p \gg l_B$ 
(see \citealt{MalyshkinKulsrud2001} for a calculation of $m_p$), the 
principal mirrors arise at scales much larger than $l_B$ and therefore 
are strong deviations of the field strength from 
the mean value found in the tail of the PDF. 

We generate a lognormal magnetic field with a Kolmogorov spectrum and the same rms value of the 
logarithm of the field strength and the same correlation length $l_B$ 
as the secularly growing mirror fluctuations analyzed in Section~
\ref{sec:mirror} (the dotted PDF in Fig.~\ref{fig:mirror_pdf}). 
The diffusivity suppression factor is 
obtained by a Monte Carlo simulation and shown in Fig.~\ref{fig:suppr} 
by the dash-dotted line. Its dependence on $\lambda/l_B$ is much steeper 
than for the mirror fields, with no constant asymptotic value at large 
$\lambda/l_B$. Qualitatively, it is quite similar to the effective 
suppression of conductivity obtained for stochastic distributions by 
\citet[see their Fig. 3]{MalyshkinKulsrud2001}.   

\subsection{Suppression of electron transport in a saturated magnetic 
field produced by MHD dynamo.}
\label{sec:mhd_sim}

We have demonstrated that the suppression of electron diffusion in a stochastic field may be 
considerably larger than in a mirror-like periodic field, most notably if $\lambda \gg l_B$. 
However, this regime is inapplicable to the magnetic fluctuations generated 
by MHD turbulence in the ICM, because there $\lambda \lesssim l_{\nu} < L_{\rmn{inj}} 
\sim l_B$, where $l_{\nu}$ is the viscous scale of turbulent eddies, 
$L_{\rmn{inj}}$ the outer (injection) scale of turbulence, and $l_B$ is 
the {\it parallel} correlation length of the magnetic field. While MHD-dynamo-produced 
magnetic fluctuations decorrelate at small (resistive) scales, it is the field's 
variation perpendicular to itself (direction reversals) that occurs at those scales, 
whereas the parallel variation is on scales $l_B$ of order the flow 
scale $L_{\rm inj} \gg \lambda$ \citep{Schek2002,SchekCowleyTaylor2004}. In the 
cool cores of galaxy clusters, $\lambda 
\sim 0.05 ~\mathrm{kpc}$, $l_{\nu}\sim 0.4 ~\mathrm{kpc}$, $L_\rmn{inj} 
\sim 10 ~\mathrm{kpc}$ (based on the parameters for the Hydra A cluster given 
by \citealt{EnsslinVogt2006}); in the hot ICM, $\lambda \sim 20 ~\mathrm{kpc}$, 
$l_{\nu}\sim 100 ~\mathrm{kpc}$, $L_{\rmn{inj}} \sim 200 ~\mathrm{kpc}$. 
Schematically, the spectrum of magnetic-field-strength fluctuations in the ICM is 
shown in Fig.~\ref{fig:Bpar_sp}. In order for magnetic trapping to be effective, 
magnetic-field-strength fluctuations $\delta B/B \sim 1$ need to 
exist at spatial scales below the electron mean free path. While mirror 
fluctuations easily satisfy this condition, MHD turbulence capable of creating 
parallel magnetic fluctuations occupies scales above $\lambda$, so large suppression 
is not expected in this case.

\begin{figure}
\includegraphics[width=84mm]{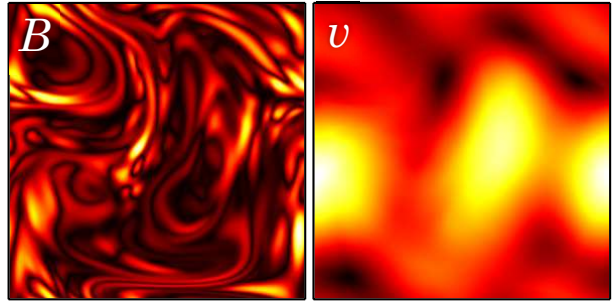}
\caption{Central snapshot of cross-sections of a dynamo-generated magnetic-field 
({\it left}) and velocity ({\it right}) magnitudes during the saturated 
turbulent MHD state for $\rmn{Pm} = 1000$, $\rmn{Re}=3$.} 
\label{fig:mhd_runs}
\end{figure}

In order to estimate an upper limit on the suppression of electron 
diffusion by MHD magnetic fluctuations, we use simulations of 
a turbulent MHD dynamo at different magnetic Prandtl numbers $\mathrm{Pm}=
\nu/\eta$, where $\nu$ is the fluid viscosity, $\eta$ the magnetic diffusivity. 
Our code solves the full set of compressible MHD equations in 3D, and is based on 
the unsplit van Leer integrator combined with the constrained transport 
method, similar to the one implemented in $ATHENA$ (\citealt{Stone2009}). 
We initiate a 3D $256^3$ periodic box of MHD plasma 
with magnetic fluctuations at the level $\beta=2000$, and stir it by a random 
white-in-time nonhelical body force applied at the largest scales 
($L_{\rmn{inj}} \sim$ the box size). As we noted earlier in this section, 
the smaller the ratio $\lambda/L_{\rmn{inj}}$ is, the less effective magnetic 
trapping is. In terms of the Reynolds number $\rmn{Re} \sim L_{\rm inj} u / 
 \lambda v_{{\rm th},i}$, it means that for small $\rmn{Ma}/\rmn{Re}$, where $\rmn{Ma}$ is the Mach 
number of the turbulent motions, conduction suppression should be negligible. 
In the cores of galaxy clusters, $\rmn{Re} \sim 100$ (Hydra A), while in 
the bulk of the ICM, $\rmn{Re} \sim 1-10$ (ignoring the possible
effects of microinstabilities on the gas viscosity). The typical 
Mach number in galaxy clusters is believed to be $\rmn{Ma}\sim0.1$ \citep[e.g.,][]
{Zhuravleva2015}. Because we seek to obtain an upper limit on suppression, 
we restrict our analysis to low $\rmn{Re}$, corresponding to the hot ICM 
(at such low $\rmn{Re}$, turbulence will not have a wide inertial range, but 
that is irrelevant because turbulent 
MHD dynamo only requires a stochastic velocity field, not necessarily a fully 
developed Kolmogorov turbulence). Namely, in our simulation, we use 
$\rmn{Re}=3$, and $\rmn{Pm}=1000$. The simulation lasts until the 
magnetic field becomes saturated: $\langle B^2 \rangle/(8\pi) \sim \langle 
\rho v^2 \rangle/2$, where $\rho$ is the mass density, $v$ is the turbulent 
plasma flow velocity (in saturation,  $\beta \sim 50$ and $\rmn{Ma} \sim 0.1$). The 
structure of the magnetic and velocity fields in the saturated state is shown 
in Fig.~\ref{fig:mhd_runs}: magnetic folds are clearly seen at the scale 
of the box, while the velocity  is stochastic but smooth, due to low $\rmn{Re}$. This 
simulation setup and the properties of the saturated magnetic field are 
similar to those of the run ``S4-sa'' in \cite{SchekCowleyTaylor2004}.

\begin{figure}
\begin{center}
\includegraphics[width=76mm]{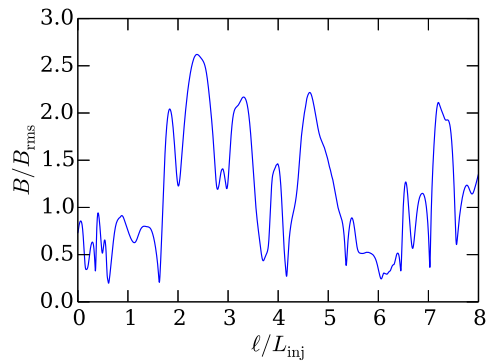}
\caption{Variation of the magnetic-field strength along a field line segment that 
spans eight boxes (box size = energy injection sale $L_{\rmn{inj}}$), taken from an MHD simulation of 
the saturated state of turbulent dynamo at $\rmn{Pm} = 1000$, $\rmn{Re}=3$.} 
\label{fig:mhd_along}
\end{center}
\end{figure}

\begin{figure}
\begin{center}
\includegraphics[width=76mm]{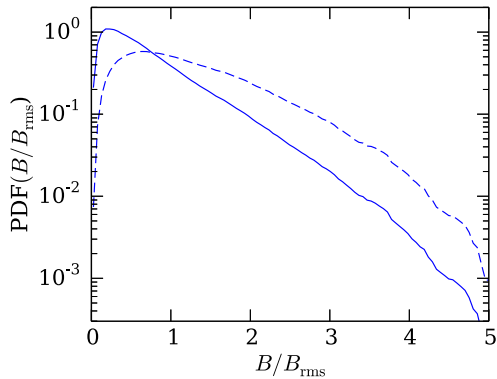}
\caption{Solid line: the 3D PDF of the magnetic-field strength in 
saturated state for $\rmn{Pm} = 1000$, $\rmn{Re}=3$. 
Dashed line: the PDF of $B$ along the field line (the 3D PDF 
multiplied by $B$).} 
\label{fig:mhd_pdf}
\end{center}
\end{figure}

\begin{figure}
\begin{center}
\includegraphics[width=76mm]{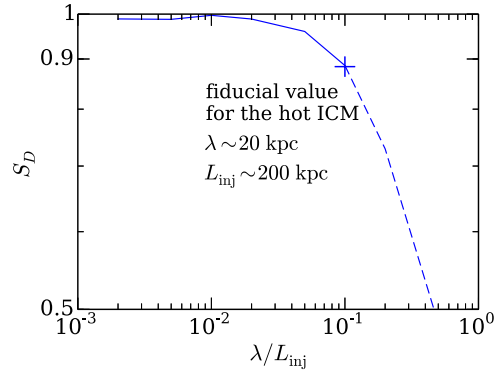}
\caption{The suppression factor $S_D=D/D_0$ of the electron diffusivity by 
turbulent-dynamo-produced magnetic fields. The cross indicates the largest 
possible suppression for the fiducial parameters of the hot ICM: $L_{\rmn{inj}} 
\sim 200~\rmn{kpc}$, $\lambda \sim 20~
\rmn{kpc}$.} 
\label{fig:mhd_suppression}
\end{center}
\end{figure}

Following the same strategy as in the case of the mirror fields, we have 
extracted a  magnetic-field line from the box in the saturated state. The 
extracted field line spans 100 box sizes (again, a field line does not 
bite its tail, although the box is periodic). This is necessary to make it 
statistically representative, because the PDF of the field now has an 
exponential tail. A line segment that spans eight boxes is shown in Fig.~
\ref{fig:mhd_along}. The PDF of the magnetic-field strength calculated 
over the whole 3D simulation domain and one calculated along the field 
(by multiplying the 3D PDF by the magnetic-field strength) 
are shown in Fig.~\ref{fig:mhd_pdf}. They clearly exhibit an exponential shape.
We calculate the suppression of electron diffusion in the same way as we did 
for the mirror fields in Section~\ref{sec:MCres}. The suppression factor 
is shown in Fig.~\ref{fig:mhd_suppression} as a function of the ratio 
of the mean free path $\lambda$ to the injection scale $L_{\rmn{inj}}$\footnote{
Note that this is a somewhat artificial parameter scan as we do not vary the 
ion mean free path, i.e., the viscosity, in a manner consistent with the 
electron mean free path.}. 
For the fiducial parameters of the hot ICM with the largest value of 
$\lambda$, we choose $L_{\rmn{inj}} \sim 200~\rmn{kpc}$ and $\lambda 
\sim 20~\rmn{kpc}$. These parameters provide maximum suppression factor of 
electron diffusion $S_D \sim 0.9$. It is shown in Fig.~\ref{fig:mhd_suppression} 
by the cross, the solid line corresponds to the suppression factor at lower $\lambda$, 
while the dashed line shows this factor for test monoenergetic 
electrons at higher $\lambda$ to better exhibit the 
shape of the function $S_D(\lambda/l_B)$ for the dynamo-generated 
magnetic field. Though in this case, there is no simple connection 
between diffusivity and thermal conductivity [like \eqref{eq:diffcond}], 
because $S_D$ now strongly depends on the mean free path (or velocity), 
the suppression of thermal conduction for $\lambda \lesssim 20~
\rmn{kpc}$ should be essentially insignificant. 

\section{Discussion}
\label{sec:disc}

It is well recognized that thermal conduction in the ICM is anisotropic in the presence 
of even an extremely weak magnetic field. A popular assumption, adopted in many theoretical 
and numerical studies, is that conduction is suppressed across the field, while 
along the field, it is equal 
to the isotropic thermal conduction in an unmagnetized plasma. This 
assumption is typically applied to large-scale fields in the ICM, e.g., to scales of order 10 kpc, 
which correspond to the characteristic field correlation length inferred from Faraday 
rotation measurements \citep[e.g.,][]{Kuchar2011}. However, the ICM is likely to be susceptible to 
microinstabilities on much smaller scales comparable with the ion Larmor radius. In 
particular, the mirror instability can generate fluctuations of the field strength of large 
amplitude $\displaystyle \delta B/B\sim 1$, which can partially suppress 
electron transport along the field lines. Given that the ion Larmor radius is some 
13 orders of magnitude smaller than the typical macroscopic scales,  
small-scale magnetic mirrors could potentially modify thermal conduction in a 
significant way, provided that mirrors can trap the electrons. 

To address this question, we have examined the properties of the field-strength fluctuations 
in the recent shearing-box simulations of the mirror instability by \cite{Kunz2014}. 
The striking difference between the magnetic fluctuations produced in these simulations 
and a generic random field is that their PDF($B)$ has sharp cutoffs at 
both low and high $B$, with the ratio of the maximal and minimal field strengths 
over the field lines $B_{\rm{max}}/B_{\rm{min}}$ $\sim$6--7. Since the ratio $B_{\rm{max}}/
B(\ell)$ determines the loss cone for a particle at the location $\ell$ along the 
line, the modest values of $B_{\rm{max}}/B_{\rm{min}}$ already suggest a limited 
amount of suppression, although it depends also on the exact shape of the PDF($B$) 
[see Section~\ref{sec:PDF}]. While we have used 2D simulations, this
imposes no obvious qualitative constraints on the mirror perturbations.
The first 3D simulations of a dynamo-generated magnetic field by \cite{Rincon2015} 
indeed appear to show qualitatively familiar-looking mirrors being generated 
along stretched field lines. While the parallel correlation length of the mirror 
fluctuations in our work is not much smaller than the size of the computational 
box, this imposes no unphysical constraints on their structure. This is because 
the box is shearing-periodic, and the scale of mirrors is set by the distance 
to marginal stability, which depends on the shear, not on the box size (see 
\citealt{Kunz2014} and \citealt{Rincon2015}).

As anticipated, in our Monte Carlo simulations, we have found that the electron diffusivity 
is suppressed by a factor of $\sim 8$ for secularly growing mirrors and 
by a factor of $\sim 5$ for saturated ones. A lognormal magnetic field with the same rms 
would produce a much stronger suppression. We further argue that the suppression of thermal 
conduction relative to an unmagnetized plasma is a factor of $\sim$2 less strong due to the 
fact that mirrors primarily affect spatial transport of the electrons, and much less 
the energy equilibration time. We conclude that microscale magnetic mirrors give rise to a 
factor of $\sim 5$ suppression of the parallel thermal conductivity. 

In this work, we assumed a static magnetic field taken from a region of a plasma 
where the field lines are stretched by a linear shear. Though at a given location in 
the ICM plasma, the field lines are not constantly stretched, the turbulent 
dynamo produces a magnetic-field-line configuration that consists of long folds 
(regions of amplified field) and short reversals (regions of decreasing/weak field). 
This means that mirror fluctuations may develop almost everywhere along the field 
lines in a turbulent ICM (see \citealt{Rincon2015} for the first numerical evidence 
of this). We note that it is not yet known how the mirror and firehose 
instabilities evolve over multiple correlation times of a turbulent velocity field. 
However, the recent results by \cite{Melville2015} indicate that at the values of 
$\beta$ typical for the ICM, the relaxation of pressure anisotropy in a changing 
macroscale velocity shear is almost insantaneous compared to the shear time. 
This may therefore suggest that the mirror instability does not have time to ever reach 
the saturated state (at $St \gtrsim 2$ according to \citealt{Kunz2014}), because the turbulent shear decorrelates 
earlier (at $St \sim 1$). Thus, secularly growing mirrors are expected to be 
more common. In any case, since the results for both phases are similar 
up to a factor of order unity, we do not expect large deviations from the 
described above behavior. We may then argue that the amount of 
suppression found using the shearing-box simulations is characteristic 
for the ICM or any other turbulent weakly collisional high-$\beta$ plasma. 

When the parallel scale of the field is larger than the particles' mean free path, the suppression 
of conductivity is not strong because collisions are frequent enough to 
stop particle trapping. This means that even though macroscopic MHD turbulence can 
produce large-scale variations of $B$, the resulting suppression of parallel conduction 
should be negligible. We illustrate this point by carrying out MHD simulations of saturated 
turbulent dynamo and explicitly calculating the suppression factor (Section~\ref{sec:mhd_sim}).

Parallel thermal conduction can also be reduced by anomalous pitch-angle scattering of 
electrons off magnetic perturbations. Such perturbations can be produced 
at the scale of the electron Larmor radius by the whistler instability triggered 
by electron pressure anisotropy \citep{Riquelme2016}. 
In the ICM, \cite{Riquelme2016} estimate the resulting effective electron mean
free path to be at most a few times smaller than the Coulomb mean free path, 
so our results remain valid (the mean free path is still much larger than the 
ion Larmor scale). The additional electron scattering will cause additional 
suppression of thermal conduction. The suppression by the mirror instability 
should then be our factor of $S_D \sim1/5 $ relative to this whistler-modified 
conductivity.

In addition to the suppression of parallel thermal conduction, the stochastic topology of the 
magnetic-field lines contributes to the total suppression of the global large-scale thermal 
conductivity by making the path travelled by an electron longer. When studying this effect, the 
effective increase of transverse diffusion due to the exponential divergence of the stochastic 
field lines should be taken into account \citep[e.g.,][]{RR1978}, because it restores the diffusive 
regime of spatial particle transport. If magnetic turbulence develops over a range of scales, the 
suppression effect is quite modest, $\sim 1/5$ of the Spitzer value \citep[][]{Narayan2001,ChandranMaron2004}. 
Since we have shown that the parallel conductivity is suppressed by another factor of $\sim 5$, 
we argue that the global large-scale thermal conduction in the ICM is roughly $\sim 1/20-1/30$ of 
the Spitzer value. \footnote{And perhaps down by another factor of a 
few if whistlers are triggered and have the effect predicted by 
\cite{Riquelme2016}.} 

\section*{Acknowledgements}

We thank S. C. Cowley for helpful discussions. The authors are grateful to W. Pauli Institute, Vienna, 
for its hospitality. E. C. acknowledges partial support by grant No. 14-22-00271 from the Russian Scientific Foundation.

\bibliographystyle{mn2e}
\bibliography{bibliography}

\appendix

\section{Transport of a passive scalar}
\label{app:A}

Assume a collisional 1D gas with a linear mean gradient of a scalar 
quantity $a$ transferred by the gas particles:
\beq 
\label{eq:ax}
\langle a(x) \rangle = {\rm const} + \alpha x.
\eeq
Here and below, the angle brackets denote averaging over the particles' 
distribution. The gradient is sustained by fixed boundary conditions (e.g., 
walls kept at constant $a$). The particles can exchange $a$ via collisions. 
Our goal is to evaluate the flux of $a$ given by 
\beq
\label{eq:qa0}
q_a = \langle a \vpar \rangle,
\eeq
where $\vpar$ is particle velocity (that is the parallel electron velocity 
along a field line in application to our problem). 

Let us first write the Langevin equation for a particle's velocity: 
\beq
\dot{v}_{\parallel} = -\nu_1 \vpar + \eta_1(t),
\eeq
where $\nu_1$ is the particle-scattering collision rate, and
$\eta_1(t)$ is a Gaussian white noise with zero mean. Solving 
for $\vpar$ gives
\beq
\vpar = \vpar(0) e^{-\nu_1 t} + \int_0^t dt' e^{\nu_1 (t'-t)} \eta_1(t'),
\eeq
and, integrating again,
\bea
\label{eq:xt}
\nonumber
x &=& x(0) + \frac{\vpar(0)}{\nu_1} (1- e^{-\nu_1 t})\\ &&+ 
	\int_0^t dt' \int_0^{t'} dt'' e^{\nu_1 (t''-t')} \eta_1(t'').
\eea
The Langevin equation for the evolution of $a$ of a given 
particle due to collisions reads:
\beq
\dot{a} = -\nu_2 (a - \langle a \rangle) + \eta_2(t), 
\eeq   
where $\nu_2$ is the $a$-exchange collision rate, and $\eta_2$ is 
a Gaussian white noise with zero mean. Solving for $a$, 
we get
\bea
\label{eq:at}
\nonumber
a &=& a(0) e^{-\nu_2 t} + 
	\nu_2 \int_0^t dt' e^{\nu_2 (t'-t)} \langle a[x(t')] \rangle \\
	&&+ \int_0^t  dt' e^{\nu_2 (t'-t)} \eta_2(t').
\eea

Combining equations~(\ref{eq:ax}), (\ref{eq:qa0}) and (\ref{eq:at}), we can 
calculate the scalar flux $q_a$ at time $t$:
\bea
\nonumber
\label{eq:qax}
q_a &=& \langle a(t) \vpar(t) \rangle = 
		\nu_2 \int_0^t dt' e^{\nu_2(t'-t)} \langle \langle a[x(t')] \rangle \vpar(t) \rangle \\
		&=& \alpha \nu_2 \int_0^t dt' e^{\nu_2(t'-t)} \langle x(t') \vpar(t) \rangle.    
\eea
The noise terms do not contribute to the flux because they all have zero mean value.
We can express $x(t')$ similar to equation~(\ref{eq:xt}) as
\bea
\nonumber
x(t') &=& x(t) - \frac{\vpar(t)}{\nu_1} [ 1-e^{\nu_1 (t'-t)}]\\
				&& + \int_t^{t'} dt'' \int_t^{t''} dt''' e^{\nu_1 (t'''-t'')} \eta_1(t''').
\eea
Substituting $x(t')$ into \eqref{eq:qax}, we get
\bea
\nonumber
q_a &=& - \alpha \langle \vpar^2(t) \rangle \frac{\nu_2}{\nu_1}  
		\int_0^t dt' e^{\nu_2(t'-t)} [ 1-e^{\nu_1 (t'-t)}] \\
		& \to & - \frac{\alpha}{3} \frac{\langle v^2 \rangle}{\nu_1 + \nu_2} 
		~\rmn{as}~t \to \infty,      
\eea
where $\langle \vpar^2(t) \rangle = (1/3) \langle v^2 \rangle$.
We see that the flux of the passive scalar $a$ is inversely 
proportional to the sum of the scattering rate of the particles 
$\nu_1$ and the $a$-exchange rate $\nu_2$. Then the 
scalar conductivity $\kappa_{a0}$ is
\beq
\label{eq:kappaa0}
\kappa_{a0} = \frac{1}{3} \frac{\langle v^2 \rangle}{\nu_1 + \nu_2}.
\eeq
If the particles only exchange $a$ and do not exchange energy, 
$\langle v^2 \rangle =v^2$.

It is also useful to derive the connection between the scalar 
flux $q_a$ and the velocity autocorrelation function. Let us 
first write $x(t')$ as
\beq
x(t') = x(t) -\int_{t'}^t \vpar(t'') dt''
\eeq
and substitute this into \eqref{eq:qax}:
\bea
\nonumber
q_a &=& - \alpha \nu_2 \int_0^t dt' e^{\nu_2(t'-t)} 
		\int_{t'}^t dt'' \langle \vpar(t'') \vpar(t) \rangle \\
\nonumber
	&=& - \alpha \nu_2 \int_0^t dt' e^{\nu_2(t'-t)} 
		\int_{t'-t}^0 d\tau \langle \vpar(t+\tau) \vpar(t) \rangle \\ 
	&\to & - \alpha \nu_2 \int_0^{\infty} dt' e^{-\nu_2 t'} \int_0^{t'} d\tau C(\tau)
	~\rmn{as}~t \to \infty,
\eea
where $C(\tau)=\langle \vpar(0) \vpar(\tau) \rangle$ is the parallel-velocity 
autocorrelation function. For the conductivity $\kappa_a$ of the scalar $a$, 
we infer
\beq
\label{eq:kappaa1}
\kappa_a = \nu_2 \int_0^{\infty} dt' e^{-\nu_2 t'} \int_0^{t'} d\tau C(\tau).
\eeq
With no magnetic mirrors, $C_0(\tau) = (1/3)v^2 e^{-\nu_1 \tau}$, and after substitution 
of $C_0$ into \eqref{eq:kappaa1}, we recover \eqref{eq:kappaa0}.  

In Section~\ref{sec:two_eff}, we demonstrated that in the limit $\lambda/l_B \gg 1$, 
the parallel velocity autocorrelation function of the monoenergetic electrons in 
the presence of mirror fluctuations has the form
\beq
C(t) = \frac{1}{3} S_p v^2 e^{-\nu_{\rm{eff}} t}.
\eeq
The coefficients $S_p$ and $\nu_{\rm{eff}}$ are determined by the Monte Carlo 
simulations. Now we can express $\kappa_a$ in terms of these two coefficients 
and the $a$-exchange rate $\nu_2$ by substituting $C(t)$ into \eqref{eq:kappaa1}:
\beq
\label{eq:kappaa2}
\kappa_a = \frac{1}{3} \frac{S_p v^2}{\nu_{\rm{eff}}+\nu_2} = \frac{1}{3} \frac{ S_p v^2}
				{(\lambda / \lambdaeff) \nu_1 +\nu_2}.
\eeq
By combining equations~(\ref{eq:kappaa0}) and (\ref{eq:kappaa2}), we 
obtain the suppression factor of the scalar conductivity $\kappa_a/\kappa_{a0}$:
\beq
\label{suppr_scalar}
\frac{\kappa_a}{\kappa_{a0}} =S_p \frac{\nu_1 + \nu_2}{(\lambda / \lambdaeff) \nu_1 +\nu_2}.
\eeq
We apply the above formula to relate the suppression of diffusion with the suppression 
of thermal conduction qualitatively, by taking $a$ to be the electron temperature.

\label{lastpage}

\end{document}